\documentstyle[pra,aps,preprint]{revtex}
\begin{document}

\author{Benjamin Paul, Michael Schulz,}
\address{Fachbereich Physik, Martin--Luther--Universit\"at Halle,
06099 Halle, Germany} 
\author{and Harry L. Frisch}
\address{Department of Chemistry, University at Albany, SUNY,
Albany, 12222 NY}
%%%%%%%%%%%%

\title{The Equation of State of a Gaussian Phantom Network with Defined Cross Link
Functionality}
\date{\today}
\maketitle
%%%%%%%%%%%%

\begin{abstract}
A phantom network of Gaussian chains far from the point of gelation can be
described as a gas of interacting particles represented by the cross links.
The type of particles varies with the network functionality, whereas the
type of interaction depends on the properties of the connecting chains. In a
mean field approximation the Equation of State can be calculated using
Mayer's cluster expansion. The resulting isothermal compressibility is
compared for different cross link functionalities.

\end{abstract}

\section{Introduction}

Even nowadays the statistical mechanics of a polymer network is far from
being fully understood. Usually a Gaussian network is described by a model
proposed by Edwards in 1968 \cite{ed68}. In his approach the cross link
constraint leads to an effective short range interaction term between free
polymer chains. This purpose is achieved by summing over all possible cross
link configurations before performing the sum of configuration of the
junctions. In fact this model is only reasonable if used in the context of
quenched cross links. Using the replica trick to determine the system's free
energy, the calculation turns out to be tedious since the cross
link--induced action term couples all replicated configuration spaces \cite
{go89}.

Still there is an approach used by H. Brodowsky and S. Prager in 1963 \cite
{pr63}, who calculated the statistical mechanics of gel formation in an
ideal phantom network. The phantom network is a model introduced by H. James
and E. Guth in 1943 \cite{jg43}, which describes the phenomenological
behaviour of a network, while assuming that the junctions may fluctuate
freely without being restricted by topological constraints due to the
existence of neighbouring chains. Prager and Brodowsky are using a
microscopic ansatz, where the network is dealt with as a gas of free cross
links interacting by means of the connecting polymer chains. It is a phantom
network, since the chains have no excluded volume term. Thus different
monomers can be placed on the same site. The chain configuration average has
been performed before averaging the cross link configuration and the
influence of the junctions is absorbed in an effective potential between
cross links. In this context the cross links are treated as being annealed.
I.e. the time of spliting up and rearranging a cross link is small compared
to typical time scales of observation. Due to the annealed limit,
topological constraints caused by entanglement of network chains are
disregarded.

\section{Calculation of the Equation of State}

In this article we demonstrate how the Brodowsky and Prager model \cite{pr63}
changes if the annealed phantom network is composed of cross links having a
predefined functionality. For clarity we assume that all connecting chains
in the network are of equal length. Returning to the general case of varying
chain length presented by Brodowsky and Prager is straightforward.

The network consists of $M$ cross links and $N$ bifunctional monomer units
forming $K$ active chains or junctions that are attached on both ends to
different cross links. Since the length of the junctions is fixed at $n$
monomer units the total number of monomers is given by $N=nK$. If only
active chains are taken into account the configurational partition function
reads \- 
$$
\zeta =\int\limits_V\cdots \int\limits_V{\rm d}{\bf r}_1\ldots {\rm d}{\bf r}%
_M\sum\nolimits^{(\sigma )}\left[ \prod_{i<j=1}^M\frac 1{\sigma
_{ij}!}\prod_{\alpha =1}^{\sigma _{ij}}q(\rho _{ij})\right]  
$$
\begin{equation}
\label{gen}\times \delta _{\sum_{i<j=1}^M\sigma
_{ij},K}\prod\limits_{i=1}^M\delta _{\sum_{j=1}^{i-1}\sigma _{ij},f}\text{ .}
\end{equation}
The function $q$ is the number of configuration that a junction $\alpha $
connecting cross link $i$ to cross link $j$ can assume . For a Gaussian
chain it is given by 
\begin{equation}
q(\rho _{ij})=\frac 3{2\pi n}\exp \left( -\frac{3\rho _{ij}^2}{2l_0^2n}%
\right) .
\end{equation}
This is the number of different configuration of a chain of $n$ units whose
ends are fixed at points a distance $\rho _{ij}=\mid {\bf r}_i-{\bf r}_j\mid 
$ apart. The elastic properties of this chain are determined by the Kuhn
length $l_{0\text{.}}$

The summation $\sum^{(\sigma )}$ extends over all ways of attaching the
chains to the $M$ available cross links. $\sigma $ is a matrix with $\sigma
_{ij}$ giving the number of junctions connecting the cross link $i$ to the
cross link $j$. Conservation of the total number of active chains is taken
into account by the first constraint $\delta _{\sum_{ij}\sigma _{ij},K}$.

The last term defines the number of chains starting from a cross link $i$.
It is given by the functionality $f$. This is assumed to be an even number
to assure the connectivity of the network. To express the constraint in a
simplified way two possible cases which can arise are distinguished. If $f$
is taken to be a single number, the Kronecker delta of the last constraint in
Eq.(\ref{gen}) can be written using Grassmann variables $\eta $ \cite{zj93}
that are defined by  
\begin{equation}
\eta _1\eta _2=-\eta _2\eta _1,\ \int \frac{{\rm d}\eta }{\sqrt{2\pi }}=0,\
\int \frac{{\rm d}\eta \eta }{\sqrt{2\pi }}=1\ {\rm and}\ \eta \eta =0\ .
\end{equation}
These properties allow us to rewrite the Kronecker delta as 
\begin{equation}
\prod_{i=1}^M\delta _{\sum_{j=1}^{i-1}\sigma _{ij},f}=\int
\prod_{i=1}^M\prod_{k=1}^f\frac{{\rm d}\eta _k^i}{\sqrt{2\pi }}%
\prod_{i<j=1}^M\left[ (\sum_{k=1}^f\eta _k^i)(\sum_{k=1}^f\eta _k^j)\right]
^{\sigma _{ij}}\ .
\end{equation}
If the functionality is abitrary but even $f=2k$ \footnote{%
NB.\ Like in the work done by Brodowsky and Prager the cross links are never
saturated. But contrary to them we assume that the functionality has to be
even at any cross link.}, the Kronecker delta can easily be expressed using
classical spins $s_i=\pm \frac 12$ 
\begin{equation}
\prod_{i=1}^M\delta _{\sum_{j=1}^{i-1}\sigma
_{ij},2k}=\sum_{\{s\}}\prod_{i<j=1}^M(s_is_j)^{\sigma _{ij}}
\end{equation}
where the sum $\sum\limits_{\{s\}}$ extends over all possible spin
configurations. For clarity the latter case will be used for our calculation
that is performed following \cite{pr63}. In a first step the partition
function 
\begin{equation}
Z=\sum_{K=0}^\infty {\rm e}^{\beta \nu K}\zeta (K)
\end{equation}
is determined. The chemical potential $\nu $ is associated to the number of
active chains $K$. The summation over $\sum_{}^{(\sigma )}$ can be performed
and yields an exponential function. Thus the partition function reads like
the partition function of a gas 
\begin{equation}
\label{gas}Z=\frac 1{M!}\sum_{\{s\}}\int\limits_V\cdots \int\limits_V{\rm d}%
{\bf r}_1\ldots {\rm d}{\bf r}_M\text{ }{\rm e}^{-\sum\limits_{i<j=1}^M\beta
s_is_j\Phi (\rho _{ij})}
\end{equation}
composed of charged particles. These are interacting via an inverted
Gaussian potential 
\begin{equation}
\Phi (\rho )=-\frac 3{2\pi n\beta }{\rm e}^\nu {\rm e}^{-(\rho /R_g)^2}
\end{equation}
where the typical length scale is given by the gyration radius of the
junctions $R_g=\sqrt{3nl_0^2/2}$. In the more detailed case of a fixed
functionality where the classical spins are replaced by Grassmann variables,
the network is mapped on a gas of particles that are composed of $f$
fermions. In the case of variable length of junctions the potential changes
to a Debye--H\"uckel kind of interaction \cite{pr63}.

The partition function of Eq.(\ref{gas}) cannot be calculated rigorously but
a virial expansion of the Equation of State can be obtained using Mayer's
cluster expansion \cite{ma37}. It is an expansion of the system's grand
partition function 
\begin{equation}
\Xi =\sum_{M=0}^\infty {\rm e}^{\mu M}Z(M)
\end{equation}
in powers of $g_{ij}=(1-\exp (-\beta s_is_j\Phi (\rho _{ij}))$. Thus the
Equation of State can be written in a series in powers of the average volume
per cross link $v=\frac VM$ 
\begin{equation}
\label{eos}\beta Pv=\left[ 1-\sum_{k=1}^\infty \frac k{k+1}\beta _k\left( 
\frac{R_g^3}v\right) ^k\right] \ .
\end{equation}

To determine the Equation of State up to the second virial coefficient the
following irreducible cluster integral has to be determined 
\begin{equation}
\beta _1=\frac{R_g^3}{ V}\int\limits_{V/R_g^3}{\rm d}\widetilde{{\bf r}}_1{\rm d}\widetilde{%
{\bf r}}_2\text{ }g_{12}\text{ }
\end{equation}
where $\widetilde{{\bf r}}={\bf r}/R_g$. The value of the first cluster
integrals for the case of abitrary but even functionality is given by

\begin{equation}
\label{first}\beta _1=\int\limits_0^\infty {\rm d}\widetilde{r}\text{ }4\pi 
\widetilde{r}^2\left[ 4-4\cosh \left( \frac \beta 2\Phi \left( \widetilde{r}%
\right) \right) \right] .
\end{equation}
The integral can only be evaluated by means of numerical methods. Using the
description for a fixed functionality $f,$ the first cluster integral
simplifies due to the reduction of the exponential function to a single
Taylor coefficient and reads%
$$
\beta _1=-\frac 4{f!}\int\limits_0^\infty {\rm d}\widetilde{r}\text{ }4\pi 
\widetilde{r}^2\left( -\frac \beta 2\Phi (\widetilde{r})\right) ^f 
$$
\begin{equation}
\label{sec}=-4\sqrt{\pi }\frac{\left( \frac 3{4\pi n}\right) ^f}{f!}%
f^{-\frac 32}\text{ .}
\end{equation}
The summation over all functionalities, starting with $f=2,$ leads to the
general result in Eq.(\ref{first}). It has to be pointed out that the
relation $\beta _1<0$ holds in both expressions of the cluster integral (\ref
{first}) and (\ref{sec}).

\section{The volume compression modulus}

Having determined the Equation of State, the volume compressibility of the
phantom network can be calculated. One has to keep in mind that the annealed
phantom network is a thermalized gas. Therefore the elastic constants that
are connected to an anisotropic deformation such as a shear modulus do not
exist. Using the Equation of State (\ref{eos}) the isothermal
compressibility i.e. the relative change of volume under isotropic
compression 
\begin{equation}
\kappa _T=-\frac 1v\left( \frac{\partial v}{\partial P}\right) 
\end{equation}
can be calculated. Expanded up to the second virial coefficient the inverse
compressibility reads 
\begin{equation}
\label{comp}\beta \kappa _T^{-1}\simeq \frac 1v-\beta _1\left( \frac{R_g^3}{%
v^2}\right) .
\end{equation}
That allows to compare three different regimes of the network.

\begin{itemize}
\item[I.] The gyration radius may be much larger then the typical distance
between cross links, given by $v^{-3}$. That is a regime where cross links
far apart are connected by a junction, $R_g^3\gg v$. Thus the connecting
chains are highly entangled so that instead of growing linear with the
volume, $\kappa _T$ is now growing with a quadratic dependence on the volume 
$V^2$. Still, due to the highly entangled junctions, there should be a
correction by the intrinsic topology of the network, that we expect to be of
the same or of higher order than the phantom network's contribution.

\item[II.] If the gyration radius is much smaller than the cross link
distance $R_g^3\ll v$, the network mainly consists of small subnetworks,
constituted of two cross links and $f$ junctions. Since the density of these
subnetworks is very small - their extension is also given by $R_g$ - the
system behaves like an ideal gas and $\beta \kappa _T^{-1}\simeq \frac 1v$.

\item[III.] It is essential to point out, that the regime, where the
assumption of a phantom network should give the main features of the
networks compressibility is within the region $R_g^3\approx v$. The inverse
compressibility is now given by an ideal gas case with a rescaled
temperature, that strongly depends on the functionality of the network (cf.
Eq. (\ref{sec}))
\end{itemize}

\begin{equation}
\frac \beta {\left( 1-\beta _1\right) }\kappa _T^{-1}\simeq \frac 1v\text{ .}
\end{equation}

These results are not comparable to the elastic equation of state, obtained
for phantom networks using a Flory approach \cite{me88}. Since the
anisotropic deformation cannot be treated by mapping the network on a gas,
only homogeneous volume change are described by the model presented in this
paper. The dependence on the functionality of the networks in the first
virial correction differs from the results of the linear compressibility
modulus due to anisotropic stress of the phantom network as given in \cite
{me88}.

Experimental or numerical results to find the volume compressibility of a
phantom network are not known to us, so that we could only speculate about
its contribution to a real polymer network. In a real system, there should
also be important contributions given by the soft-- (or hard--) core
potentials of the molecules, as well as contributions due to entanglement
effects. Still, the model of Brodowsky and Prager allows to find the volume
compressibility using a very general and straightforward scheme.

\section{Acknowledgments}

This work has been partially supported by the Deutsche
Forschungsgemeinschaft schu 934/1-3, NSF Grant DMR 962 8224 and the Donors
of the Petroleum Fund of the American Chemical Society.

\end{document}